\documentclass[preprint,showpacs,preprintnumbers,amsmath,amssymb,aps,pre]{revtex4-2}
\usepackage{amsmath,amssymb}
\usepackage{epsfig}
\usepackage{graphics}
\usepackage{graphicx}
\usepackage{color}

\begin{document}

\title{Stochastic birhythmicity outside the coexistence region in the Hindmarsh-Rose model}

\author{Ignacio Ortega-Piwonka}
\email[]{juanignacio.ortega@urjc.es}
\affiliation{Nonlinear Dynamics, Chaos and Complex Systems Group, Departamento de
F\'{i}sica, Universidad Rey Juan Carlos, Tulip\'{a}n s/n, 28933 M\'{o}stoles, Madrid, Spain}

\author{Javier Used}
\affiliation{Nonlinear Dynamics, Chaos and Complex Systems Group, Departamento de
F\'{i}sica, Universidad Rey Juan Carlos, Tulip\'{a}n s/n, 28933 M\'{o}stoles, Madrid, Spain}

\author{Jes\'{u}s M. Seoane}
\affiliation{Nonlinear Dynamics, Chaos and Complex Systems Group, Departamento de
F\'{i}sica, Universidad Rey Juan Carlos, Tulip\'{a}n s/n, 28933 M\'{o}stoles, Madrid, Spain}

\author{Miguel A.F. Sanju\'{a}n}
\affiliation{Nonlinear Dynamics, Chaos and Complex Systems Group, Departamento de
F\'{i}sica, Universidad Rey Juan Carlos, Tulip\'{a}n s/n, 28933 M\'{o}stoles, Madrid, Spain}

\date{\today}


\begin{abstract}
In this work, we demonstrate that the Hindmarsh-Rose model subjected to additive white noise exhibits birhythmicity. Specifically, the system fluctuates between two distinct bursting attractors characterized by different numbers of spikes. This behavior is observed not only within the bistable region bounded by two saddle-node bifurcations of limit cycles but also beyond these boundaries. This phenomenon is associated with the ghost effect, typically observed near deterministic saddle-node bifurcations. We map the region of stochastic birhythmicity in terms of the noise intensity and a key deterministic parameter that controls the dynamics of fast ion channels. To provide an analytical foundation, we introduce a \textcolor{black}{simple} stochastic model with a single saddle-node bifurcation. In this \textcolor{black}{model}, stochastic birhythmicity is similarly characterized as a function of noise intensity and the control parameter.
\end{abstract}
\maketitle

\newpage

\section{Introduction} \label{sec_Introduction}

The human brain is one of the most complex dynamical systems in nature. A proper understanding of such an intricate structure requires in first place comprehension of its fundamental units: the neurons. Over the last century, numerous models based on differential equations and difference equations have been proposed to this end. One of the most relevant of these models was introduced in 1952 by Alan Hodgkin and Andrew Huxley, inspired by their study on the squid giant axon. This model describes the initiation and propagation of action potentials in neurons, which are mathematically understood as excitable systems \cite{Hodgkin_Huxley, Izhikevich}. In 1984, J. L. Hindmarsh and R. M. Rose proposed a new model focused on another phenomenon experimentally observed in individual neurons: \textit{bursting}, a periodic regime consisting of sequences of rapid firing of spikes occurring between periods of quiescence \cite{ Hindmarsh_Rose_ProcRoySoc_1984}. One of the most interesting phenomenon observed in oscillatory dynamical systems like neurons is the fluctuations between different stable regimes. This phenomenon, known as multirhythmicity, has taken great interest in the last years and it refers to the coexistence of two or more limit cycles describing different types of oscillations \cite{Goldbeter_2022}. In nonlinear birhythmic or multirhythmic systems, the presence of random disturbances can cause dramatic changes in the dynamical behavior of the system. This effect is specially significant when the stochastic trajectories are close to the separatrices that delimit different basins of attraction and produce mixed-mode oscillatory regimes \cite{Nganso_2022}.

In this paper, we explore the parameter space of the \textcolor{black}{stochastic} Hindmarsh-Rose neuron model. We demonstrate that, under the influence of noise, the region in which the system exhibits birhythmicity extends beyond the bistable region predicted in the deterministic case, which is bounded by two saddle-node bifurcations. This extension is attributed to the ghost effect, a phenomenon commonly associated with systems near such bifurcations. The stochastic birhythmicity is numerically characterized as a function of the noise intensity and a key parameter governing the dynamics of fast ion channels in the model. \textcolor{black}{An analytical or probabilistic study of the stochastic Hindmarsh-Rose model is challenging due to its lack of symmetry. Instead, we propose a simple, axially symmetric stochastic model that reproduces the aforementioned phenomenon, allowing for a characterization based on its associated Fokker-Planck equation.}

The organization of this paper is as follows. In Sec.~\ref{sec_HR_model}, we provide a detailed description of the Hindmarsh-Rose model under the influence of additive white Gaussian noise, highlighting its impact on the system's birhythmic behavior. A \textcolor{black}{simple} model featuring a saddle-node bifurcation of a periodic solution is introduced in Sec.~\ref{sec_LCF}. We show that, under the influence of noise, random transitions occur between two states on either side of the saddle-node, resembling the stochastic birhythmicity observed in the Hindmarsh-Rose model. Finally, the main conclusions of this work are summarized in Sec.~\ref{sec_conclusions}.

\section{Hindmarsh-Rose model} \label{sec_HR_model}

The Hindmarsh-Rose model is a system of nonlinear differential equations that reproduces behavior typically observed in individual biological neurons regarding firing patterns, slow-fast dynamics and bursting \cite{ Hindmarsh_Rose_ProcRoySoc_1984, Barrio_Chaos_2020_HR_homoclinic, Slepukhina_et_al_Chaos_2023},
  \begin{align}\label{eq_Hindmarsh_Rose}
    \begin{split}
      \dot{x} &= y - ax^3 + bx^2 -z + I,\\
      \dot{y} &= c - dx^2 - y,\\
      \dot{z} &= r(s( x -x_0) - z) + \varepsilon\xi(t).
    \end{split}
  \end{align}

Here, $x$ represents the potential difference across the neuron's membrane, while $y$ and $z$ are the \textcolor{black}{fast} and \textcolor{black}{slow} variables, respectively, governing the activation and inactivation of ion currents through the membrane channels. The normalized parameters $a,b,c$, and $d$ regulate the dynamics of the fast ion channels, whereas $r$\textcolor{black}{, $s$ and $x_0$} control the slow channels. \textcolor{black}{Specifically, these parameters tune the shape of the nullclines, in order to reproduce phenomena observed in the more complex Hodgkin-Huxley model \cite{Hindmarsh_Rose_ProcRoySoc_1984}. $I$ is the current injected into the neuron.} The term $\varepsilon\xi(t)$ captures the intrinsic randomness in the neuron, arising from various microscale factors not accounted for by the deterministic variables. Specifically, $\varepsilon$ is the noise intensity, while $\xi(t)$ is a normalized, time-uncorrelated additive white noise function, satisfying the following equations,
  \begin{align}\label{eq_white_noise}
    \begin{split}
      \langle\xi(t)       \rangle &= 0,\\
      \langle\xi(t)\xi(t')\rangle &= \delta(t'-t),
    \end{split}
  \end{align}
where $\delta(t'-t)$ is the Dirac delta function. \textcolor{black}{Following Ref. \cite{Slepukhina_et_al_Chaos_2023}, the noise is applied on the slow gating variable $z$ only, since random fluctuations in this variable is what essentially leads to transitions between stable attractors. Adding noise in the $x$ and $y$ equations would have a minor effect on this regard.}

In this work, the parameter $b$ is varied between $2.9$ and $2.93$, while $\varepsilon$ is varied between $10^{-5}$ and $10^{-2}$. \textcolor{black}{Following Ref. \cite{Slepukhina_et_al_Chaos_2023}, t}he other parameters are fixed at $a=1$, $c=1$, $d=5$, $s=4$, $x_0=-1.6$, $r=0.01$, and $I=2.2$. \textcolor{black}{These fixed values are recurrent in several works and reproduce output data collected in experiments involving pond snail neurons \cite{Hindmarsh_Rose_ProcRoySoc_1984, Slepukhina_et_al_Chaos_2023, Barrio_Chaos_2020_HR_homoclinic}.} For these fixed values and varying $b$, the deterministic model ($\varepsilon=0$) exhibits a branch of bursting periodic solutions. As $b$ decreases, the branch folds back and forth in saddle-node bifurcations, each time increasing the number of bursting spikes in just one spike \cite{Slepukhina_et_al_Chaos_2023, Barrio_Chaos_2020_HR_homoclinic}. \textcolor{black}{For fixed parameter values other than those mentioned above, this behavior is qualitatively similar. A thorough study on this matter is available in \cite{Barrio_Chaos_2020_HR_homoclinic}.}

\begin{figure}
    \centering\includegraphics[width=\linewidth]{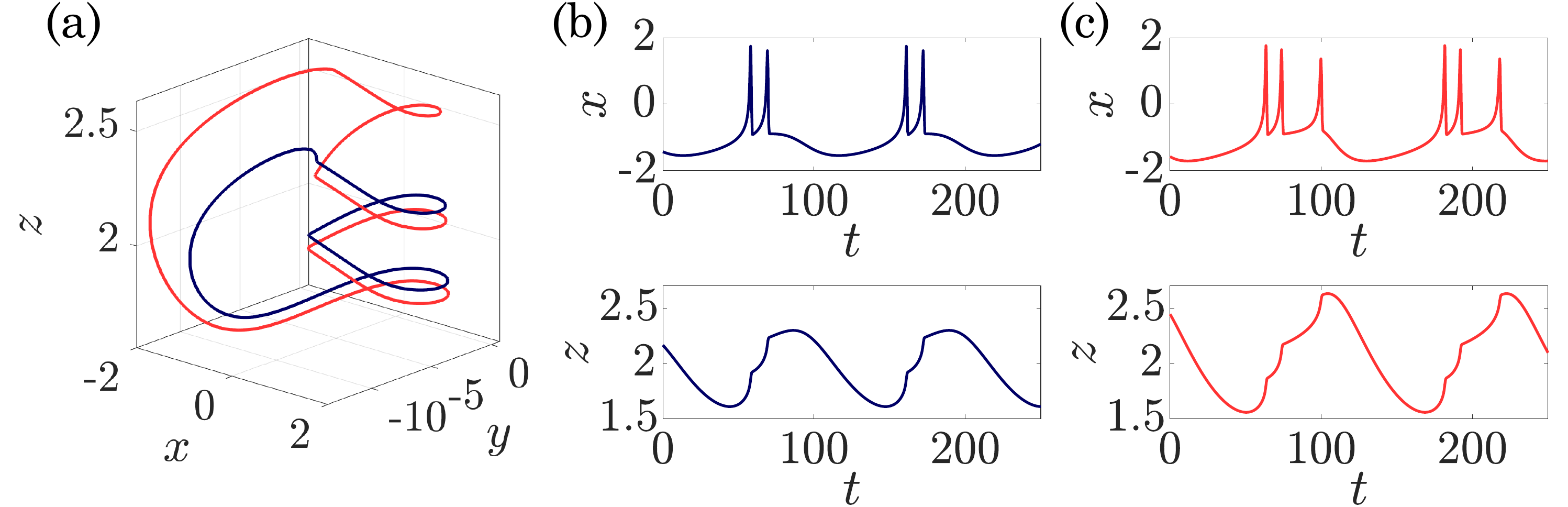}
    \caption{Stable periodic bursting solutions of the Hindmarsh-Rose model, Eq.~(\ref{eq_Hindmarsh_Rose}), for $b=2.916$ and $\varepsilon=0$. (a) Time series of the 2-cycle (blue) and 3-cycle (red) orbits in the phase space. (b) $x$ and $z$ coordinates of the 2-cycle as a function of time. (c) $x$ and $z$ coordinates of the 3-cycle as a function of time.}
    \label{fig_birhythmicity}
\end{figure}

Subsequently, two distinct stable periodic bursting solutions coexist between consecutive saddle-node bifurcations, leading to bistability in the system. This coexistence of two stable periodic solutions is referred to as \textit{birhythmicity} \cite{Slepukhina_et_al_Chaos_2023}. Two of these saddle-node bifurcations take place at about $b=2.9082$ and $b=2.9231$. For $b>2.9082$, a stable periodic solution with \textcolor{black}{$2$} spikes  emerges, while for $b<2.9231$, a periodic solution with \textcolor{black}{$3$} spikes appears. We refer to these solutions as the \textcolor{black}{$2$}-cycle and \textcolor{black}{$3$}-cycle, respectively. Within the interval $2.9082 \leq b \leq 2.9231$, both the 2-cycle and 3-cycle coexist, defining a region of deterministic birhythmicity. Figure~\ref{fig_birhythmicity} illustrates examples of the 2-cycle and 3-cycle solutions at a single point within this region. In both cases, the oscillation period is divided into a slow, quiescent phase, where the variables change at a steady pace, and a fast, bursting phase, where the variables $x$ and $y$ oscillate quickly and thus emit spikes. The $z$ variable is always slow and does not show spikes.

\begin{figure}
    \centering\includegraphics[width=0.7\linewidth]{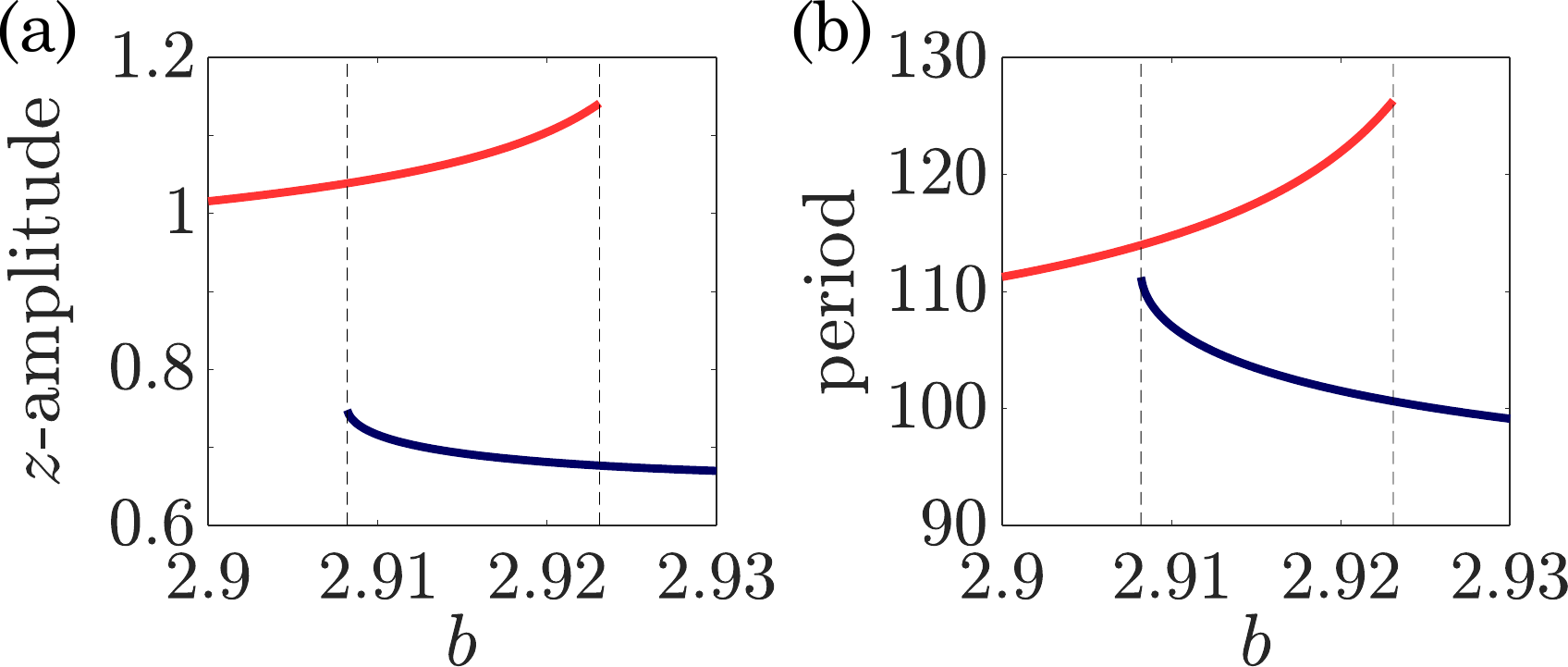}
    \caption{Numerical computation of the bifurcation diagram of the deterministic Hindmarsh-\textcolor{black}{Rose} model ($\varepsilon=0$), showing the $z$-amplitude (a) and the period (b) of the stable periodic bursting solutions, namely, 2-cycle (blue) and 3-cycle (red). At $b=2.9082$ and $b=2.9231$, the two stable branches fold in saddle-node bifurcations. Therefore, the model exhibits birhythmicity when $2.9082 \leq b \leq 2.9231$.}
    \label{fig_det_bif_diag}
\end{figure}

Figure~\ref{fig_det_bif_diag} summarizes the evolution of both the 2-cycle and 3-cycle branches as the control parameter $b$ varies. Specifically, the cycle \textcolor{black}{peak-to-peak} amplitude along the $z$-axis \textcolor{black}{(from now on, the $z$-amplitude)} and the period are shown in Fig.~\ref{fig_det_bif_diag}(a) and Fig.~\ref{fig_det_bif_diag}(b), respectively. As $b$ increases, the $z$-amplitude and period of the $3$-cycle increase, while those of the 2-cycle decrease.

\begin{figure}[t]
    \centering\includegraphics[width=0.7\linewidth]{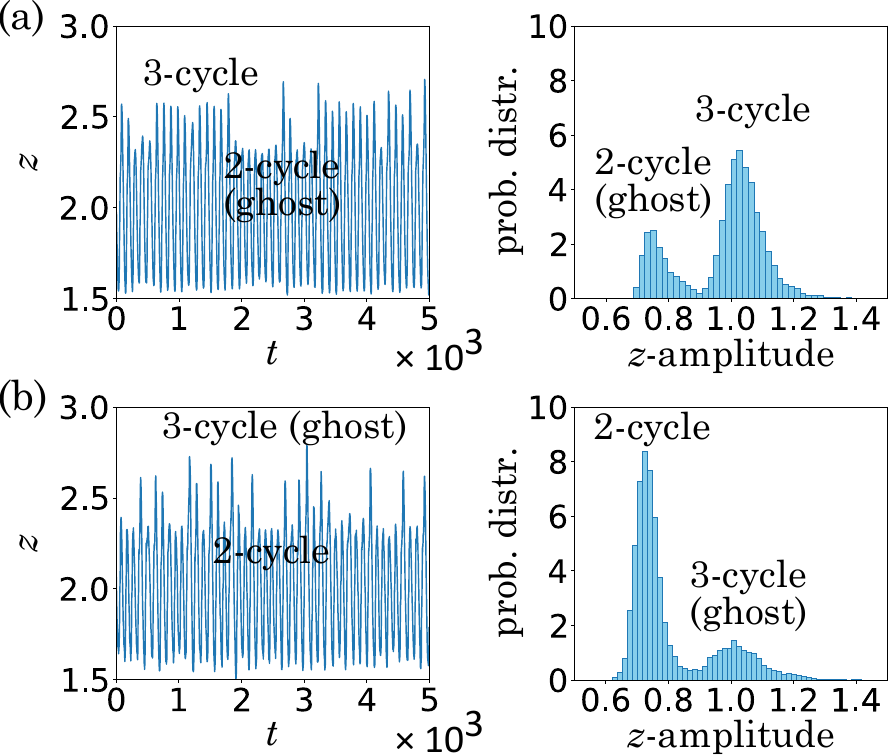}
    \caption{Numerical simulations of the stochastic Hindmarsh-Rose model for different parameter values: (a) $b=2.907, \varepsilon=4\times10^{-3}$. (b) $b=2.924, \varepsilon=8\times10^{-3}$. The time series of the $z$ variable are shown as well as normalized histograms of the $z$-amplitudes estimated in each oscillation. As in both cases the system lies outside, but close to the birhythmicity region, the injected noise triggers transitions between the unique bursting periodic attractor and the ghost of the other bursting periodic attractor.}
    \label{fig_ghost_stochastic}
\end{figure}

The effects of incorporating noise ($\varepsilon>0$) into the Hindmarsh-Rose model in the deterministic birhythmicity region have already been reported in \cite{Slepukhina_et_al_Chaos_2023}. Here, the random external driving induces transitions between the two coexisting stable periodic solutions, as it randomly drives the system across the separatrix that delimits their basins of attraction. In this work, we report that the random transitions between \textcolor{black}{the} bursting regimes mentioned above also take place outside the deterministic birhythmicity region. The phenomenon is more prominent as the noise intensity is increased. Figure~\ref{fig_ghost_stochastic} summarizes the dynamics of the model when white noise is injected, for values of $b$ outside the deterministic birhythmicity region but close to the saddle-node bifurcations. \textcolor{black}{Specifically, the slow variable $z$ is observed because it does not exhibit bursting, which makes it easier to study numerically and measure its amplitude in each oscillation. In addition, the $z$-amplitude is well correlated with the number of peaks per burst in $x$ and $y$ (see Figs.~\ref{fig_birhythmicity} and \ref{fig_det_bif_diag}). Indeed, bursts with two peaks have a $z$-amplitude below $0.9$, while bursts with three peaks have a $z$-amplitude above $0.9$.} In Fig.~\ref{fig_ghost_stochastic}(a), the control parameter $b$ is set at the \textcolor{black}{left} side of the birhythmicity region, where only the 3-cycle attractor exists, but the initial condition is set close to the 2-cycle attractor. In Fig.~\ref{fig_ghost_stochastic}(b), the parameter $b$ is set at the \textcolor{black}{right} side, where only the 2-cycle attractor exists, with an initial condition close to where the 3-cycle would be expected to appear. In both cases, the system randomly transitions between the 2-cycle and the 3-cycle, similar to the behavior observed within the birhythmicity region. This is evident in the $z$ time series, which shows a sequence of small and large amplitude oscillations occurring randomly, corresponding to the amplitudes of the 2-cycle and 3-cycle, respectively, as illustrated in Fig.~\ref{fig_birhythmicity}. This behavior is further highlighted in the histogram of the $z$-amplitude, where two distinct peaks are clearly visible.

\begin{figure}[t]
    \centering\includegraphics[width=0.9\linewidth]{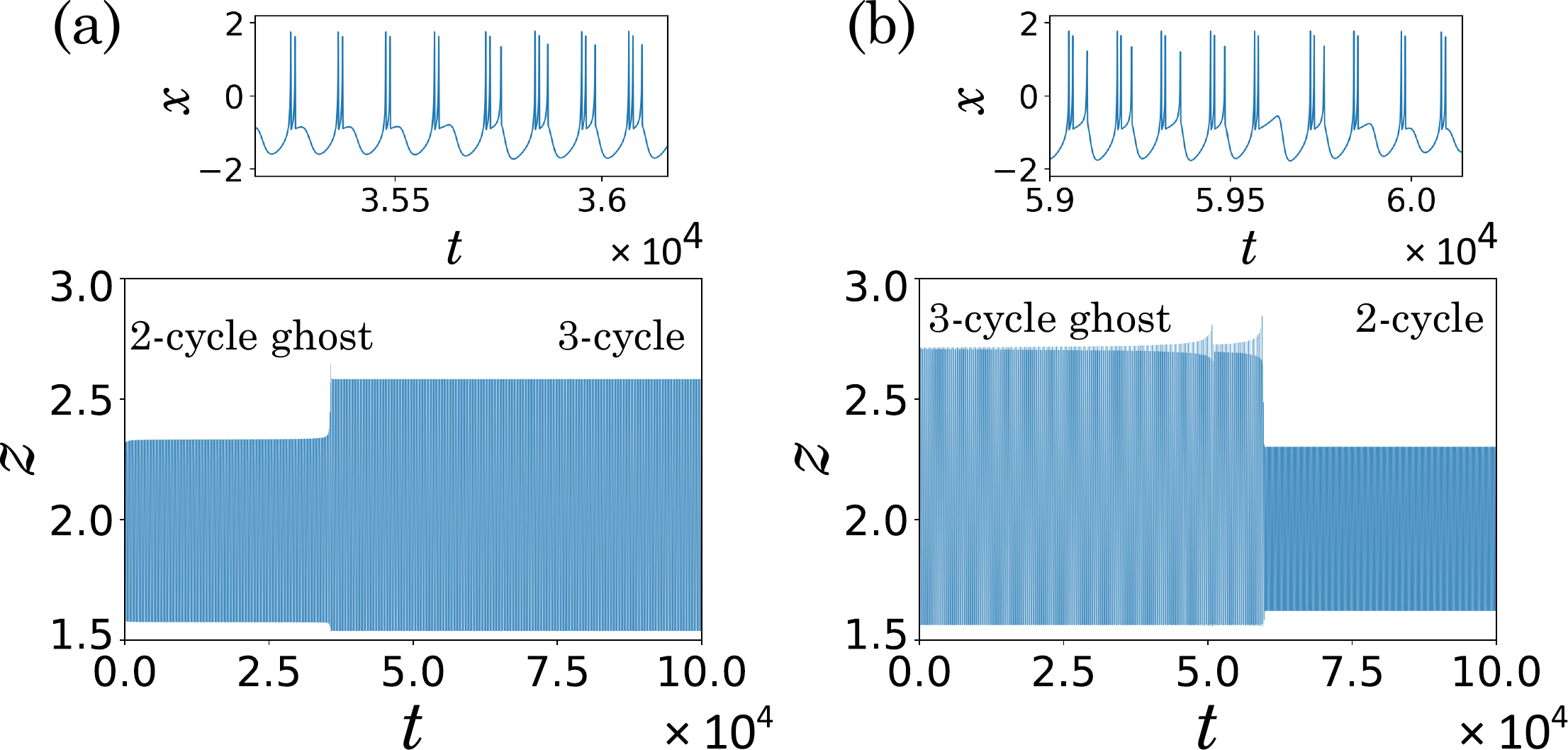}
    \caption{Simulations of the deterministic Hindmarsh-Rose model ($\varepsilon=0$) for different values of $b$ and initial conditions: (a) $b=2.90816$, with initial conditions \textcolor{black}{close to} the $2$-cycle \textcolor{black}{ghost}. (b) $b=2.92315$, with initial conditions \textcolor{black}{close to} the $3$-cycle \textcolor{black}{ghost}. In both cases, the model exhibits a unique deterministic attractor, but given the proximity to the deterministic birhythmicity region, the time series remains close to \textcolor{black}{the} ghost for a long time before converging to the actual attractor. \textcolor{black}{As the transition occurs, the $z$-amplitude changes abruptly. The miniatures show the variable $x(t)$, zoomed in around the transition, where the change in the number of peaks per burst can be appreciated.}}
    \label{fig_ghost_deterministic}
\end{figure}

The random transitions between the $2$-cycle and $3$-cycle states in the stochastic Hindmarsh-Rose model, observed beyond the deterministic birhythmicity region, are due to the \textit{ghost} phenomenon, which is typical of deterministic systems exhibiting a saddle-node bifurcation \cite{Izhikevich}. In the proximity to the bifurcation, where no stable or unstable equilibrium exist, simulations with proper initial conditions seem to converge to the stable equilibrium because they slow down where it is expected to arise. However, the system eventually accelerates and diverts from here. The closer the system is to the saddle-node bifurcation, the longer the ghost is expected to last before it drifts away. Indeed, this phenomenon occurs in the simplest saddle-node normal form \cite{Izhikevich}. The deterministic Hindmarsh-Rose model also exhibits the ghost phenomenon in the vicinity of the deterministic birhythmicity region. Figure~\ref{fig_ghost_deterministic} illustrates numerical simulations of Eq.~(\ref{eq_Hindmarsh_Rose}) with $\varepsilon=0$ and values of $b$ outside the deterministic birhythmicity region, but close to the saddle-node bifurcations. In Fig.~\ref{fig_ghost_deterministic}(a), $b=2.90816$, the control parameter is close to the birhythmicity region at the left side, where only the $3$-cycle attractor exists, but the initial condition is set close to the $2$-cycle attractor. The system remains for a long time orbiting in what seems to be a stable bursting regime with two peaks, but eventually, after a long transient time, it quickly transitions into the $3$-cycle. Likewise, in Fig.~\ref{fig_ghost_deterministic}(b), $b=2.92315$, which means the system is also outside the birhythmicity region but at the other side. Here, only the $2$-cycle attractor exists, but with an initial condition close to the orbit of the $3$-cycle attractor, the system will follow bursting orbits with three peaks for a long time and then it will quickly jump into the $2$-cycle attractor.

When noise is introduced into the model and $b$ lies within the deterministic birhythmicity region, the system transitions between the two stable attractors, the $2$-cycle and the $3$-cycle. However, when $b$ lies outside this region but close to it, the system still transitions between the now unique stable attractor and the ghost of the other attractor. In all cases, the system exhibits stochastic birhythmicity, without distinguishing between actual attractors and their ghosts. As a result, the noise effectively extends the birhythmicity region.

\begin{figure}[t]
    \centering\includegraphics[width=\linewidth]{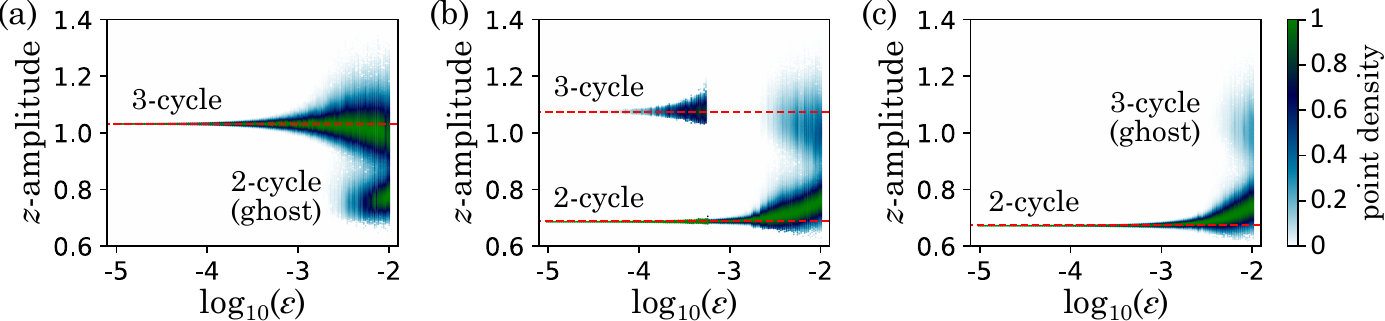}
    \caption{Scatter plots of the $z$-amplitude as a function of $\varepsilon$, computed from simulations of the stochastic Hindmarsh-Rose model for different values of $b$: (a) $b=2.906$. (b) $b=2.916$. (c) $b=2.924$. \textcolor{black}{For each $(b,\varepsilon)$, two simulations are run with initial conditions in the 2-cycle orbit and the 3-cycle orbit, respectively. About $1000$ oscillations are observed per simulation.} The diagrams show individual $z$-amplitudes computed in each oscillation as points, with colors ranging from white to blue to green according to increasing density of points \textcolor{black}{(the density of points is given in arbitrary units, with its maximal value set as 1)}. The red dashing lines correspond to the deterministic $z$-amplitudes of the stable cycles. Outside the deterministic birhythmicity region, stochastic ghosts of the $2$-cycle and $3$-cycle arise under sufficiently high $\varepsilon$.}
    \label{fig_rand_bif_diag}
\end{figure}

Figure~\ref{fig_rand_bif_diag} provides a qualitative approach onto how the noise gives rise to transitions between the 2-cycle and the 3-cycle. For different values of the parameters $b$ and $\varepsilon$, simulations of Eqs.~(\ref{eq_Hindmarsh_Rose}--\ref{eq_white_noise}) are run and, after a reasonably long transient, the $z$-amplitudes of individual orbits are registered. \textcolor{black}{As explained before, the $z$-amplitude can be used to determine if the system is in the 2-cycle or the 3-cycle and thus, statistical data about the $z$-amplitude can be used to estimate probabilities for the system to be in either cycle.} In Fig.~\ref{fig_rand_bif_diag}(a), the control parameter $b$ is set slightly at the left side of the deterministic birhythmicity region. Here, the $3$-cycle is the only stable attractor. However, with enough noise introduced into the neuron, about $\varepsilon=3\times10^{-3}$ or more, the system randomly transitions between the $3$-cycle and the $2$-cycle ghost to the point that, for $\varepsilon > 6\times10^{-3}$, both states have more-or-less equal probability to be occupied, with amplitudes that fluctuate greatly from their average values.

In Fig.~\ref{fig_rand_bif_diag}(b), the control parameter $b$ is in the middle of the deterministic birhythmicity region, and both the $2$-cycle and $3$-cycle are stable attractors. If the noise introduced is low, the system keeps orbiting in one of the cycles depending on the initial condition, with little probability to transition to the other cycle. For $\varepsilon$ between $6\times10^{-4}$ and $3\times10^{-3}$, the system remains mostly orbiting in the $2$-cycle while the $3$-cycle seems to vanish. This occurs presumably because the $2$-cycle is a much less energetic state than the $3$-cycle and thus the system is more susceptible to be kicked by the noise from the $3$-cycle to the $2$-cycle than the other way around \cite{Izhikevich, Strogatz}. For $\varepsilon$ over $3\times10^{-3}$, the $3$-cycle arises again since the noise is now so intense that is able to move the system between the two attractors in both directions regardless of the energy difference. This rough energy-based approach is valid even for a non-conservative model such as Hindmarsh-Rose, since the right-hand side of any differential equation can always be written as $-\nabla U + \nabla \times \mathbf{A}$, where $U$ and $\mathbf{A}$ are a scalar field an\textcolor{black}{d} a vector field, respectively \cite{Helmholtz_decomposition}. \textcolor{black}{A more precise, quantitative energy-based analysis would require to construct the fields $U$ and $\mathbf{A}$ from the right hand side of Eq.~(\ref{eq_Hindmarsh_Rose}). However, such an analysis would fall outside the scope of this study.}

In Fig.~\ref{fig_rand_bif_diag}(c), the parameter $b$ is set slightly at the right side of the deterministic birhythmicity region. According to the deterministic model, only the $2$-cycle attractor exists here. However, if the noise intensity is sufficiently high ($\varepsilon = 5\times 10^{-3}$ or more), the system can randomly transition to the $3$-cycle ghost and be found in that orbit with a significant probability, although still smaller than that of the $2$-cycle.

\begin{figure}[t]
    \centering\includegraphics[width=0.6\linewidth]{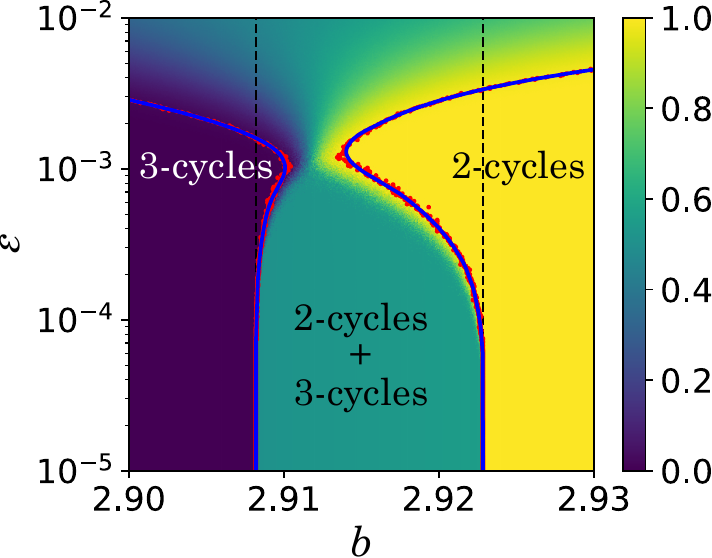}
    \caption{Proportion of $2$-cycle oscillations out of all oscillations observed in simulations of the stochastic Hindmarsh-Rose model run on the $(b,\varepsilon)$ parameter space. For each ($b,\varepsilon$), $10$ simulations with initial condition close to the $2$-cycle and $10$ simulations with initial condition close to the $3$-cycle are carried out, with a total of about $20000$ oscillations being observed. The black dashed lines mark the saddle-node bifurcations seen for $\varepsilon=0$ and thus delimit the deterministic birhythmicity region. For given values of $\varepsilon$, the red dots mark the values of $b$ between which both $2$-cycles and $3$-cycles occur with a proportion of $2\%$ or more. The blue curves are splines of the red dots. Hence \textcolor{black}{the central} region constitutes the region where stochastic birhythmicity is numerically observed \textcolor{black}{(cyan)}. For a sufficiently high level of noise, the stochastic birhythmicity region extends beyond the deterministic birhythmicity region and ghost-like orbits of the $2$-cycle and $3$-cycle are recurrently triggered by the noise.}
    \label{fig_HR_space_of_param}
\end{figure}

A more complete and quantitative perspective on how the noise affects the emergence of ghost-like orbits is provided in Fig.~\ref{fig_HR_space_of_param}. For each point in the $(b,\varepsilon)$ parameter space, simulations of Eqs.~(\ref{eq_Hindmarsh_Rose}--\ref{eq_white_noise}) with initial conditions in the $2$-cycle and $3$-cycle in equal number are run. After a reasonably long transient, $2$-cycle and $3$-cycle individual orbits are counted. The proportion of $2$-cycle orbits out of the total number of orbits is illustrated in the color map. The figure includes a numerical estimation of the boundaries \textcolor{black}{(blue lines)} of the central region \textcolor{black}{(cyan)} where both $2$-cycle and $3$-cycle orbits occur with a proportion of at least $2\%$ each\textcolor{black}{. Thus, this region} is regarded as a \textit{de facto} region where stochastic birhythmicity occurs. This does not necessarily mean that outside this region only one cycle has $100\%$ probability to occur, but one of the cycles is orbited much more \textcolor{black}{often} than the other \textcolor{black}{(all in all, the 2\% proportion was chosen as a rather arbitrary cutoff)}.

\textcolor{black}{Fig.~\ref{fig_HR_space_of_param} illustrates how the stochastic birhythmicity region narrows before it widens as $\varepsilon$ increases. For $\varepsilon = 3\times 10^{-4}$, this region is delimited by the deterministic saddle-node bifurcations. As $\varepsilon$ increases, transitions between the $2$-cycle and $3$-cycle orbits start to occur. However, close to the deterministic saddle-node bifurcations, one of the orbits is either a ghost or a newly formed, high-energy attractor, and transitions towards the other, low-energy attractor are more likely to occur than the reverse. Thus, the low-energy attractor is predominantly observed, which narrows the stochastic birhythmicity region.} There is a bottleneck at about $\varepsilon = 10^{-3}$, where the region becomes the narrowest, occupying only a third of the deterministic birhythmicity region. For higher $\varepsilon$, the stochastic birhythmicity region \textcolor{black}{widens} again, since now transitions in both directions occur \textcolor{black}{quite often}. The left boundary of the stochastic birhythmicity region trespasses the left saddle-node bifurcation at about $\varepsilon = 2\times 10^{-3}$, and the right boundary trespasses the right saddle-node bifurcation at about $\varepsilon = 3\times 10^{-3}$. In consequence, the stochastic birhythmicity region becomes wider than the deterministic one. It is here where the ghost-like orbits are most likely to emerge.

Finally, within most of the interior of the stochastic birhythmicity region, both $2$-cycle and $3$-cycle orbits occur with the same \textcolor{black}{proportion}. Below the bottleneck this happens because the system is unlikely to transition from the attractor set at initially. Above the bottleneck however, \textcolor{black}{where the system is mostly driven by the noise,} the opposite happens, and the system transitions between both attractors \textcolor{black}{often} and in both directions.

\section{Stochastic birhythmicity in a simple model} \label{sec_LCF}

The stochastic birhythmicity phenomenon is not exclusive of the Hindmarsh-Rose model. \textcolor{black}{However, a characterization of the phenomenon based on an analytical study of either the stochastic equations or the associated Fokker-Planck equation proves to be difficult due to their high level of asymmetry.} In this section, we introduce a simple, normalized, \textcolor{black}{axially symmetric} two-variable model that exhibits a periodic solution saddle-node bifurcation. It is demonstrated that, when additive noise is applied, the model shows a similar stochastic birhythmicity phenomenon, where the system randomly transitions between two attractors (or their ghosts), even beyond the region of deterministic coexistence. The simplicity of this model allows for an analytical explanation of the stochastic birhythmicity phenomenon through a probabilistic analysis. The model is governed by the following vector equation,
  \begin{equation}\label{eq_LCF}
    \frac{d\mathbf{r}}{dt} = \mathbf{F}(\mathbf{r}) + \varepsilon \boldsymbol{\xi}(t).
  \end{equation}

Here, $\mathbf{r}=\big(x(t),y(t)\big)$ encompasses the two variables. In the right-hand side of Eq.~(\ref{eq_LCF}), a deterministic field and a stochastic function are recognized. The deterministic field $\mathbf{F}(\mathbf{r})$ is in turn split into a radial part and a vortex part,
  \begin{equation}\label{eq_F_f}
    \mathbf{F}(\mathbf{r}) = f(r)\mathbf{\hat{r}} + r\omega(r)\boldsymbol{\hat\theta},
  \end{equation}
where the radial part is given by $f(r) = -r\big( (r^2-1)^2 - b \big)$ and $r=|\mathbf{r}|=\sqrt{x^2+y^2}$. \textcolor{black}{$\mathbf{\hat{r}}$ and $\boldsymbol{\hat\theta}$ are the unit vectors of polar coordinates.} The function $f(r)$ can also be written as the negative gradient of a scalar potential, \textcolor{black}{i.e., $f(r)=-du/dr$, where} $u(r) = \tfrac{1}{2} \big( \tfrac{1}{3}(r^2-1)^3 - br^2 \big)$. \textcolor{black}{For the rest of this paper, t}he vortex part is left as constant and normalized, $\omega(r)=1$. \textcolor{black}{The stochastic function} $\boldsymbol{\xi}(t)$ is a two-variable, normalized, uncorrelated additive white noise function,
  \begin{align}\label{eq_white_noise_2D}
    \begin{split}
      \langle \xi_i(t)          \rangle &= 0, \\
      \langle \xi_i(t) \xi_j(t')\rangle &= \delta_{ij}\delta(t'-t),
    \end{split}
  \end{align}
where $\delta_{ij}$ is the Kronecker delta and $\delta(t'-t)$ is the Dirac delta function. \textcolor{black}{It can be demonstrated that $\boldsymbol{\xi}(t)$ also has axial symmetry, i.e., its probabilistic behavior is invariant under rotations \cite{Gardiner}.} The parameters $b$ and $\varepsilon$ fulfill the same role as those used in the Hindmarsh-Rose model in Sec.~\ref{sec_HR_model}. The parameter $b$ acts as the control parameter that defines a deterministic region of coexistence between two stable states while $\varepsilon$ accounts for the noise intensity.

\begin{figure}[t]
    \centering\includegraphics[width=0.6\linewidth]{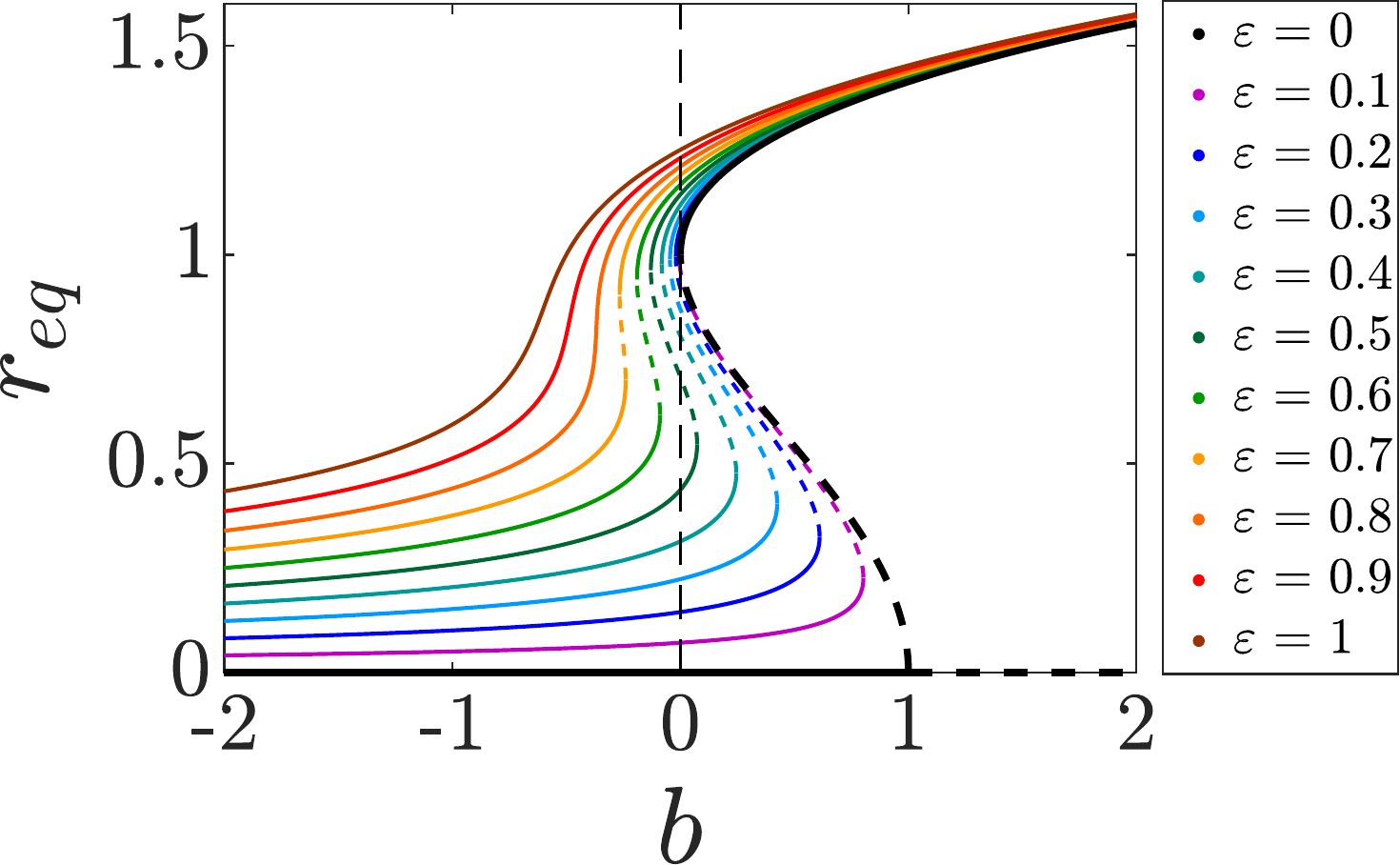}
    \caption{Bifurcation diagram of the stochastic system given by Eqs.~(\ref{eq_LCF}--\ref{eq_white_noise_2D}). The curves are given by Eq.~(\ref{eq_loc_m}) and correspond to the values of $r$ that locally maximize (solid curves) and minimize (dashed curves) the stationary probability distribution $p_s(r)$ in terms of the control parameter $b$. The colors correspond to different noise intensities $\varepsilon$. The black curves account for the stable (solid) and unstable (dashed) radii of equilibrium solutions in the deterministic case ($\varepsilon=0$).}
   \label{fig_LCF_bif_diag}
\end{figure}

The deterministic counterpart ($\varepsilon=0$) of the \textcolor{black}{simple} model in polar coordinates reads, $\dot{r}=f(r)$, $\dot{\theta}=\omega$. It is well known that this deterministic model exhibits a periodic solution saddle-node bifurcation \cite{Strogatz}. Indeed, all the equilibrium states are either the fixed point at the origin ($\mathbf{r}=\mathbf{0}$) or a circumferential limit cycle centered at the origin with angular velocity $\omega$ and radius $r$ given by $f(r)=0$. Besides $r=0$, the latter equation may have up to two solutions, depending on $b$. The solution $r=\sqrt{1+\sqrt{b}}$ exists only for $b \geq 0$ and corresponds to a stable limit cycle, while the solution $r=\sqrt{1-\sqrt{b}}$ exists only for $b\in[0,1]$ and corresponds to an unstable limit cycle. The stability of the solutions is determined by an analysis on the second derivatives of $u(r)$. The equilibrium solution branches in the deterministic case are showcased in Fig.~\ref{fig_LCF_bif_diag} (black curves). For $b<0$, the only equilibrium solution is the fixed point $\mathbf{r}=\mathbf{0}$, and it is stable. At $b=0$, the stable and unstable limit cycles emerge at $r=1$ from a saddle-node bifurcation. As $b$ increases, the stable limit cycle increases in size while the unstable limit cycle shrinks, until it coalesces with the origin at $b=1$ in a subcritical Andronov-Hopf bifurcation, and thus the origin becomes an unstable equilibrium point. Consequently, the deterministic \textcolor{black}{simple} model exhibits a bistability region at $0 \leq b \leq 1$, where the attractor at the origin and the stable limit cycle coexist.

When noise is introduced into the \textcolor{black}{model} ($\varepsilon>0$), concepts such as equilibrium and stability do not make sense anymore, at least as how they are understood in deterministic systems. Instead, a statistical description is required, based on probability distributions of possible states \cite{Gardiner, Agez_Clerc_Louvergneaux_PRE_2008, Agez_Clerc_Louvergneaux_Rojas_PRE_2013, Ortega_Clerc_Falcon_Mujica_PRE_2010}. The notion of stable and unstable equilibrium states is replaced by an new notion of states where the system is \textit{likely} to remain close to or to drift away from. These stable and unstable stochastic equilibrium states locally maximize or minimize a probability distribution that statistically describes the state of the system in the steady stage, after a sufficiently long time such that the initial condition, or its uncertainty, does not influence the random evolution of the system. The probability distribution $\rho(\mathbf{r},t)$ of the stochastic process $\mathbf{r}(t)$ ruled by Eq.~(\ref{eq_LCF}) evolves over time following the Fokker-Planck equation \cite{Fokker_FPeq_1914, Planck_FPeq_1917, Kolmogorov_FPeq_1931},
  \begin{equation}\label{eq_Fokker_Planck}
    \frac{\partial\rho}{\partial t} = -\nabla\cdot\mathbf{F}(\mathbf{r}) + \tfrac{1}{2}\varepsilon^2\nabla^2\rho.
  \end{equation}

In the steady stage mentioned above, the probability distribution is stationary, i.e., it does not have time dependence. Given the \textcolor{black}{axial symmetry} of the system, it does not depend on the angular coordinate $\theta$ either. Thus, the stationary Fokker-Planck equation reads,
  \begin{equation}\label{eq_Fokker_Planck_s}
    0 = \frac{1}{r} \frac{d}{dr} \bigg( r \Big( -f(r)\rho_s + \frac{\varepsilon^2}{2}\frac{d\rho_s}{dr} \Big) \bigg).
  \end{equation}

In the search for a stationary probability distribution with non-diverging integration over the whole two-dimensional plane, the above equation can be reduced to simply,
  \begin{equation}\label{eq_Fokker_Planck_s2}
    \frac{\varepsilon^2}{2} \frac{d\rho_s}{dr} + \frac{du}{dr}\rho_s=0,
  \end{equation}
since $f(r)=-du/dr$. The solution of Eq.~(\ref{eq_Fokker_Planck_s2}) is found to be \cite{Gardiner, Agez_Clerc_Louvergneaux_PRE_2008, Ortega_Clerc_Falcon_Mujica_PRE_2010},

  \begin{equation}\label{eq_rho_s}
    \rho_s(r) = Ce^{-\frac{2}{\varepsilon^2}u(r)},
  \end{equation}
where $C$ is a normalization constant. This is the stationary, \textcolor{black}{axially symmetric} probability distribution of the two-dimensional state $\mathbf{r}$. The probability distribution of the scalar distance $r$ to the origin is found by integrating $\rho_s$ over the circumference of radius $r$,
  \begin{equation}\label{eq_p_s}
    p_s(r) = 2\pi C r e^{-\frac{2}{\varepsilon^2}u(r)}.
  \end{equation}

While the deterministic system exhibits a limit cycle that orbits around the origin, the stochastic system exhibits stochastic equilibrium states where there is a tendency to orbit around the origin at a fluctuating distance. The most likely values of this distance to be measured at any moment are those that locally maximize $p_s(r)$. Likewise, there are distances where it is unlikely to find the system orbiting at, which locally minimize $p_s(r)$. These stochastic states are analogous to the stable and unstable deterministic equilibrium states that locally minimize or maximize a potential energy in conservative systems. This allows to extrapolate concepts such as equilibrium states, stability, branches and bifurcations to stochastic systems, and to provide quantitative descriptions in terms of the distances mentioned above. The stochastic equilibrium states are given by the equation $dp_s/dr=0$, which in our \textcolor{black}{simple} model reduces to the following,
  \begin{equation}\label{eq_loc_m}
    b = -\frac{\varepsilon^2}{2r^2} + (r^2-1)^2.
  \end{equation}

Equation~(\ref{eq_loc_m}) can be written as a third-degree polynomial equation for $r^2$ and therefore it may exhibit up to three positive solutions depending on $b$ and $\varepsilon$. Alternatively, for fixed $\varepsilon$, its current form provides a characterization of the stochastic equilibrium branch as a uniquely evaluated function $b(r)$. Some of these curves for $\varepsilon$ between $0.1$ and $1$ are included in Fig.~\ref{fig_LCF_bif_diag}. For low $\varepsilon$, the function $b(r)$ has an N-shaped profile, with a region of negative derivative embedded in between two regions of positive derivative, delimited by a local maximum and a local minimum. Also, the curve remains close to the deterministic limit cycle branches and the deterministic stable fixed point branch at $r=0$. As $\varepsilon$ increases, the stochastic equilibrium branch moves away from the deterministic branches and the aforementioned local minimum and maximum become closer to one another and, eventually, they coalesce and vanish. For even higher $\varepsilon$, the stochastic equilibrium branch has positive derivative everywhere. As $b$ diverges to $-\infty$, the stochastic equilibrium branch decays and, as $b$ diverges to $+\infty$, it gets asymptotically closer to the deterministic stable limit cycle branch.

\begin{figure}[t]
    \centering\includegraphics[width=0.6\linewidth]{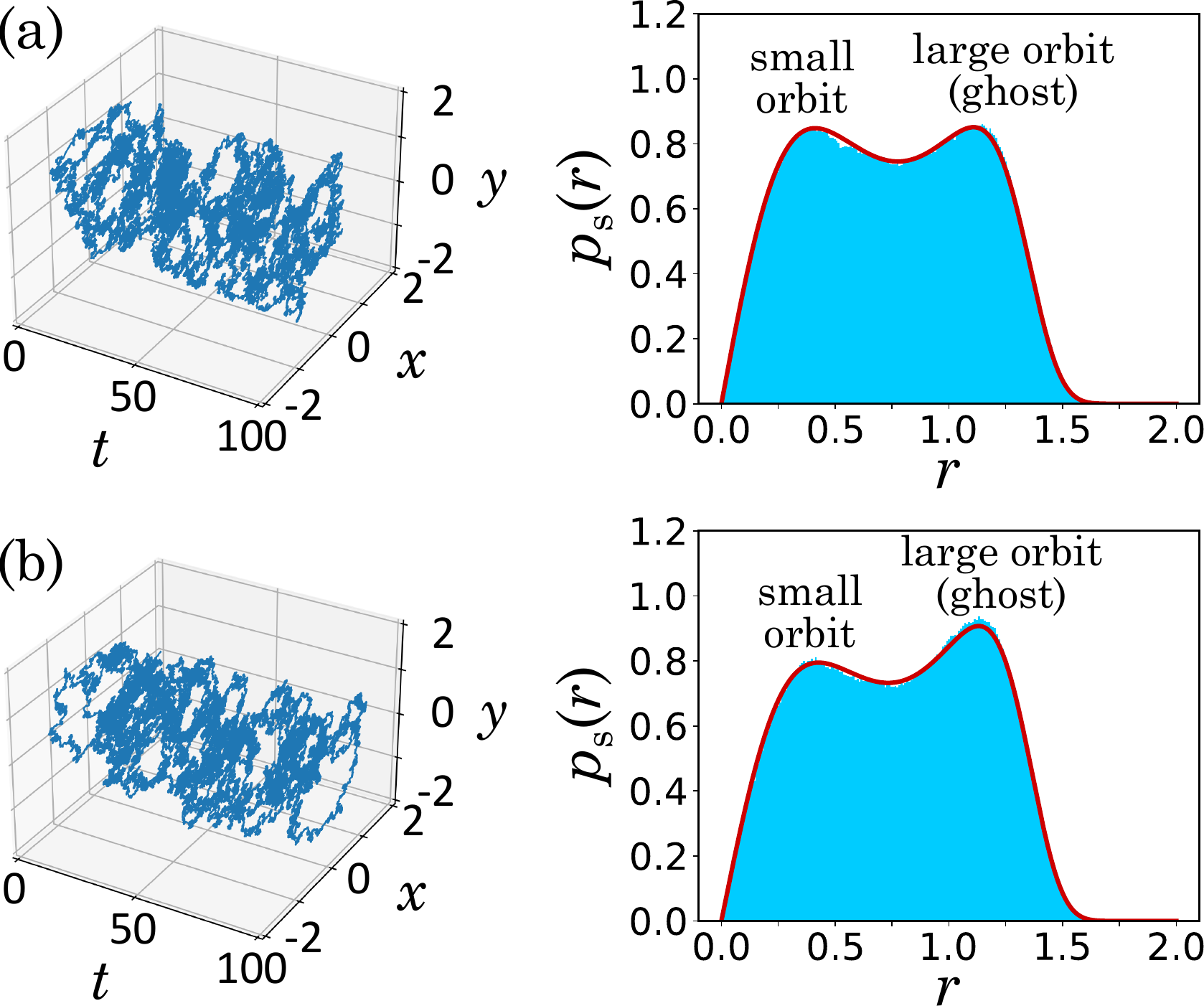}
    \caption{Numerical simulations of Eqs.~(\ref{eq_LCF}--\ref{eq_white_noise_2D}) for \textcolor{black}{$\varepsilon=0.5$} and different values of $b$. (a) $b=-0.05$. (b) $b=-0.02$. The coordinates $x$ and $y$ as function of time from a single simulation are showcased, as well as normalized histograms of $r=\sqrt{x^2+y^2}$ from $200$ simulations with disperse initial conditions. Two stochastic stable states that tend to orbit around the origin can be recognized in each case. A small one, which stems from the deterministic fixed point attractor $\mathbf{r}=\mathbf{0}$ and a large one, that arises as a consequence of the ghost that precedes the deterministic stable limit cycle. Each histogram includes the theoretically predicted probability distribution, Eq.~\textcolor{black}{(}\ref{eq_p_s}\textcolor{black}{)}, shown as a solid red line.}
   \label{fig_LCF_example}
\end{figure}

As explained before, a stochastic equilibrium state is stable (unstable) if it maximizes (minimizes) $p_s(r)$, i.e., if the second derivative of $p_s(r)$ is negative (positive). It can be demonstrated that, at the equilibrium points,
  \begin{equation}
    \frac{d^2 p\textcolor{black}{_s}}{d r^2}\bigg|_{r=r_\text{eq}} = -\frac{4\pi C}{\varepsilon^2} (r_\text{eq})^{2} \frac{db}{dr}\bigg|_{r=r_\text{eq}} e^{-\frac{2}{\varepsilon^2}u(r_\text{eq})}.
  \end{equation}

Here, $r_\text{eq}$ is any real positive solution of Eq.~(\ref{eq_loc_m}), and $db/dr$ is the derivative of the function $b(r)$ given by Eq.~(\ref{eq_loc_m}). This shows that the regions of the stochastic equilibrium branch given by Eq.~(\ref{eq_loc_m}) with positive derivative correspond to stable states, while the region with negative derivative, if present, corresponds to an unstable state. Thus, the branch folds are \textcolor{black}{stochastic} saddle-node bifurcations. The stable branch section below the saddle-node bifurcations is the stochastic counterpart of the deterministic attractor at $\mathbf{r}=\mathbf{0}$, while the stable branch section above the saddle-node bifurcations is the stochastic counterpart of the deterministic stable limit cycle. The unstable branch section in between the saddle-nodes is the stochastic counterpart of the deterministic unstable limit cycle. \textcolor{black}{Between the stochastic saddle-node bifurcations}, both stochastic stable states coexist and the system exhibits stochastic birhythmicity, i.e., both states have a probability to be observed at a certain moment. Figure~\ref{fig_LCF_bif_diag} shows that as the noise intensity increases, the saddle-nodes move closer together, eventually coalescing. As a result, the stable branches merge, and the unstable branch disappears.

Figure~\ref{fig_LCF_bif_diag} also shows that, for sufficiently high $\varepsilon$, part of the stable branch section that stems from the  deterministic stable limit cycle exists for negative values of $b$, where the deterministic system does not exhibit any limit cycle. This is a consequence of both the noise and the ghost phenomenon. A couple of examples are shown in Fig.~\ref{fig_LCF_example}. Here, simulations of Eqs.~(\ref{eq_LCF}--\ref{eq_white_noise_2D}) are run for $b<0$ and $\varepsilon>0$. At first sight, the system seems to remain close to the origin and circulate around it at a fluctuating distance. However, when the normalized histogram of such distance is computed, we can see that there are two most likely values that locally maximize the probability distribution. The smallest of these values corresponds \textcolor{black}{to} the stable branch that stems from the deterministic attractor at $\mathbf{r}=\mathbf{0}$, while the largest value corresponds to the stable branch that stems from the deterministic limit cycle. This state exists beyond the deterministic saddle-node bifurcation because the noise drives the system into the ghost that precedes the deterministic stable limit cycle at random times, similar to what happens in the stochastic Hindmarsh-Rose model.

\begin{figure}[t]
    \centering\includegraphics[width=\linewidth]{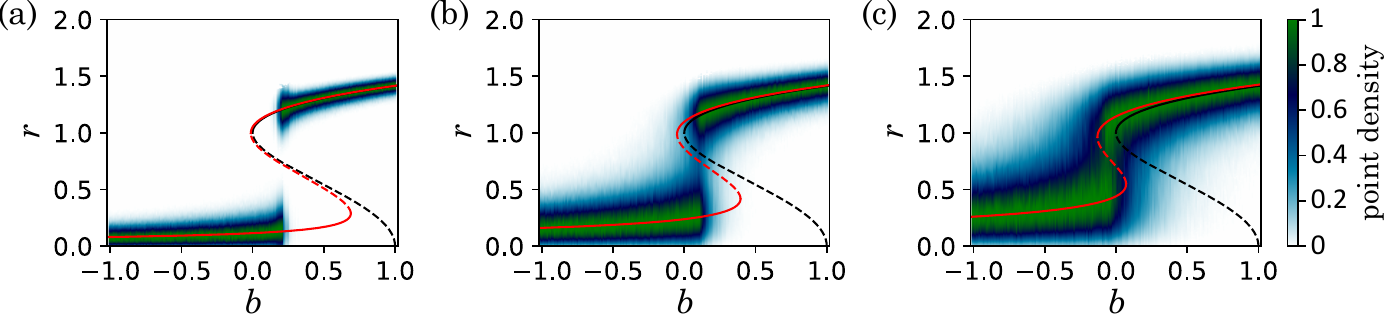}
    \caption{Scatter plots of $r$ versus $b$, computed from simulations of Eqs.~(\ref{eq_LCF}--\ref{eq_white_noise_2D}) \textcolor{black}{run} for different values of $\varepsilon$. (a) $\varepsilon=0.158$. (b) $\varepsilon=0.316$. (c) $\varepsilon=0.511$. Individual values of $r$ are depicted as points, with colors ranging from white to blue to green according to increasing density of points \textcolor{black}{(the density of points is given in arbitrary units, with its maximal value set as 1)}. The black lines account for the deterministic limit cycle branches, while the red lines account for the stochastic equilibrium branches. Solid and dashed lines account for stable and unstable states, respectively.}
   \label{fig_LCF_rand_bif_diag}
\end{figure}

Figure~\ref{fig_LCF_rand_bif_diag} provides some more statistics about the dynamics of the \textcolor{black}{simple model} around the stochastic equilibrium states discussed above. As expected, $r$ fluctuates around the most likely values that locally maximize $p_s(r)$, shown by solid red curves, given by Eq.~(\ref{eq_loc_m}). These fluctuations become wider in range as the noise intensity increases, to the point where the intervals of confidence around different coexisting local maxima seem to merge. However, there seems to be a disagreement between the theoretical prediction and numerical simulations in the sense that, for relatively low noise levels, the ranges where the \textcolor{black}{model} exhibits stochastic birhythmicity are seemingly much narrower than predicted by the theory. This apparent disagreement occurs because, just like in the stochastic Hindmarsh-Rose model, one of the two coexisting stable states may have a much higher probability to be observed and transitions triggered by noise are much more likely to occur towards it rather than from it. \textcolor{black}{Longer simulations would allow for the less likely stable states to be eventually observed, thus matching the theoretically predicted bifurcation diagrams. On the other hand}, with enough noise intensity, jumps in both directions become more frequent and the probabilities to observe both states become more balanced.

\begin{figure}[t]
    \centering\includegraphics[width=0.6\linewidth]{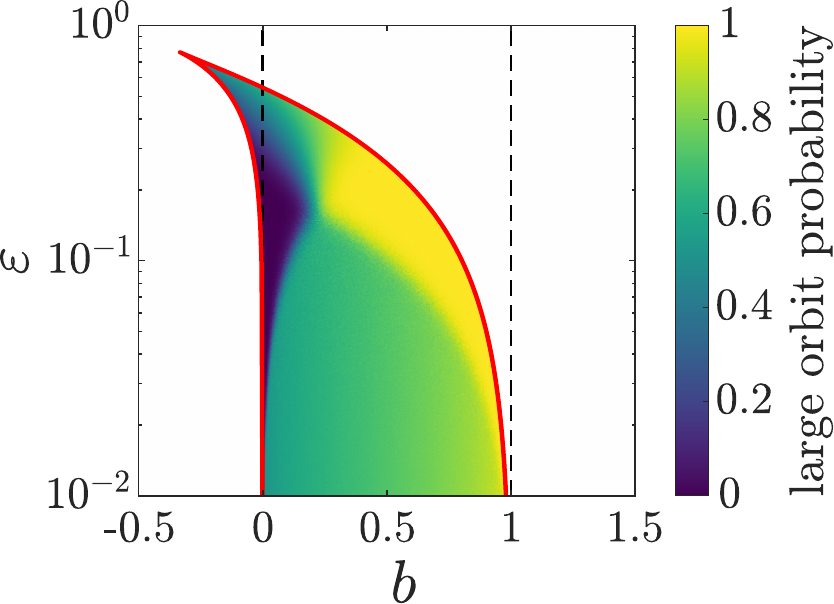}
    \caption{Numerical estimation of the probability of the \textcolor{black}{simple model} to be found in the large orbit \textcolor{black}{stochastic state}, as a function of $(b,\varepsilon)$. This probability can only be computed in between the stochastic saddle-nodes (solid red line), as anywhere else there is only one stochastic equilibrium state. The stochastic saddle-node branches emerge from the deterministic saddle-node and Hopf bifurcations, which delimit the deterministic bistability region (dashed black lines). The stochastic saddle-node branches coalesce in a cusp at the upper-left corner of the figure. For sufficiently high $\varepsilon$, the stochastic birhythmicity region trespasses the deterministic bistability region as a result of the confluence between the noise and the ghost phenomenon.}
    \label{fig_LCF_space_of_param}
\end{figure}

A more descriptive landscape of the probabilistic behavior of the stochastic \textcolor{black}{simple model} in the birhythmicity region is provided by Fig.~\ref{fig_LCF_space_of_param}. Here, the probability of the large orbit stochastic state being observed at any moment, is depicted on the $(b,\varepsilon)$ plane. The stochastic birhythmicity region has a pointy shape, with a cusp located at its top. Inside, three zones can be identified in terms of the probability distribution. In the purple zone at the left side, the small orbit stochastic state is much more dominant and, although theoretically the large state still has a non-zero probability of being observed, it seldomly appears in numerical simulations. In the yellow zone at the right side, the whole opposite happens; the large state has an almost unitary probability to be seen. As explained before, these two zones exist because, although both regimes have a non-zero probability to occur, one \textcolor{black}{has a} significantly \textcolor{black}{higher probability to occur} than the other, and there is a privileged direction for jumps between both regimes to be triggered by the noise. In the green zone at the middle, the probabilities of both states are quite balanced, with a tendency for the large state to be increasingly more likely to be observed as $b$ increases. This zone has a rather hourglass shape, with the bottleneck being located at about $(b,\varepsilon)=(0.25,0.16)$. All these characteristics discussed above are shared by the birhythmicity region in the stochastic Hindmarsh-Rose model (Fig.~\ref{fig_HR_space_of_param}).

The stochastic saddle-node branches that delimit the region of stochastic birhythmicity can be analytically characterized by considering that these bifurcations take place at the folds of the profile $b(r)$ from Eq.~(\ref{eq_loc_m}), i.e., its derivative becomes zero,
  \begin{equation}\label{eq_saddle_node}
    \frac{db}{dr} = \frac{\varepsilon^2}{r^3} + 4r(r^2-1) = 0.
  \end{equation}

From Eq.~(\ref{eq_saddle_node}), an expression for the noise intensity is obtained, $\varepsilon = 2r^2\sqrt{1-r^2}$. This expression, together with Eq.~(\ref{eq_loc_m}), where $\varepsilon$ is expanded, provide a parametric form for the sadle-node branches in the $(b,\varepsilon)$ plane, as a function of $r$,
  \begin{align}\label{eq_saddle_node_parametric}
    \begin{split}
      b           &= (1-r^2)(1-3r^2),\\
      \varepsilon &= 2r^2\sqrt{1-r^2}.
    \end{split}
  \end{align}

The stochastic saddle-node branches are illustrated in Fig.~\ref{fig_LCF_space_of_param} using solid red curves. These branches emerge from the deterministic saddle-node and Hopf bifurcations at the points $(b,\varepsilon)=(0,0)$ and $(b,\varepsilon)=(1,0)$, respectively. The branches become closer as $\varepsilon$ increases and coalesce at the cusp of the stochastic birhythmicity region. This cusp can be considered the stochastic counterpart of a pitchfork bifurcation because here, the three stochastic equilibrium states that exist in the birhythmicity region merge in a single stable stochastic equilibrium state.

The cusp location in the parameter space is given by, $(b,\varepsilon) = \big( -1/3 ,\, 4/(3\sqrt{3}) \big)$. We know this because at this point, the local maximum and local minimum of the profile $b(r)$ from Eq.~(\ref{eq_loc_m}) that define the stochastic saddle-nodes, coalesce and become a single inflection point. Thus, not only the first derivative of $b(r)$ is zero in the cusp, but the second derivative is zero as well,
  \begin{equation}\label{eq_pitchfork}
    \frac{d^2b}{dr^2} = -3\frac{\varepsilon^2}{r^4} + 4(3r^2-1) = 0.
  \end{equation}

Equations~(\ref{eq_loc_m}, \ref{eq_saddle_node}, \ref{eq_pitchfork}) constitute a system of equations, and its solution is given by, ${r=\sqrt{2/3}}$, ${b=-1/3}$, ${\varepsilon=4/(3\sqrt{3})}$. The cusp is a singular point in the sense that the saddle-node branch does not have a class C$^1$ parametric form with a moving tangent vector that does not become zero at the cusp, which is indeed the case with Eq.~(\ref{eq_saddle_node_parametric}). Equivalently, the saddle-node branch does not have a class C$^1$ \textit{arclength} parametric form because such a parametric form cannot be differentiable at the cusp.

\section{Conclusions}
 \label{sec_conclusions}
In this work, we have numerically analyzed the Hindmarsh-Rose model subjected to additive white noise. We observed that the model exhibits birhythmicity beyond the limits of the bistable region predicted by the deterministic model, which is bounded by two saddle-node bifurcations of periodic solutions. This phenomenon results from the interplay between noise and the ghost phenomenon, typically observed near saddle-nodes. To better understand this behavior, the region of stochastic birhythmicity has been characterized in terms of \textcolor{black}{the} noise intensity and the parameter $b$, which is associated with the fast ion channels. Probability distributions for the $2$-cycle and $3$-cycle orbits have been computed for different values of these parameters. The boundaries between regions where only the $2$-cycle or the $3$-cycle orbit occur, and the region where both cycles coexist, have been estimated numerically.

To provide an analytical explanation for the stochastic birhythmicity phenomenon, we propose a \textcolor{black}{simple} stochastic model with a single periodic solution saddle-node bifurcation. This model exhibits a similar phenomenon, which has been statistically characterized in terms of \textcolor{black}{the} noise intensity and a control parameter that emulates the corresponding parameter in the Hindmarsh-Rose model. By analyzing the probability distribution of the scalar distance $r$ to the origin, we are able to extrapolate concepts such as equilibrium, stability, state branches, and bifurcations to stochastic systems, providing quantitative descriptions in terms of $r$. As a result of the interplay between noise and the ghost phenomenon, stochastic states resembling the stable deterministic limit cycle emerge beyond the periodic solution saddle-node bifurcation. The saddle-node branches delimiting the stochastic birhythmicity region have been demonstrated both numerically and analytically.

\textcolor{black}{It is worth noting that, while the Hindmarsh-Rose model exhibits two stable limit cycle attractors with a birhythmicity region between two saddle-node bifurcations, the deterministic simple model exhibits a limit cycle and a steady state, with their region of coexistence delimited by a saddle-node bifurcation and a Hopf bifurcation. Still, the stochastic attractors that arise from both regimes in the presence of noise are somewhat similar to a limit cycle, in the sense that the system tends to circulate around the origin at a fluctuating distance. This tendency to circulate is described by the vortex term $\omega(r)$ in Eq.~(\ref{eq_F_f}). Overall, there is a significant difference with the stochastic Hindmarsh-Rose model. When noise is injected into the simple model, the stochastic birhythmicity region extends beyond the deterministic saddle-node bifurcation but not beyond the Hopf bifurcation, since the latter does not exhibit ghost attractors in its vicinity.}

\section*{ACKNOWLEDGMENTS}
 This work has been supported by the Spanish State Research Agency (AEI) and the European Regional Development Fund (ERDF, EU) under Projects No.~PID2019-105554GB-I0 (MCIN/AEI/10.13039/501100011033) and No.~PID2023-148160NB-I00 (MCIN/AEI/ 10.13039/501100011033).


\end{document}